\documentclass[9pt,a4paper,twocolumn,twoside]{tau-class/tau}
\usepackage[english]{babel}

\journalname{Managing Financial Risk}
\title{Comparative Evaluation of VaR Models: Historical Simulation, GARCH-Based Monte Carlo, and Filtered Historical Simulation}

\author[1]{Tian Xin}
\affil[1]{Carey Business School, Johns Hopkins University, Baltimore, MD, USA\\ \texttt{xtian21@jh.edu}}
\affil[2]{Lab for Data Science, University of Science and Technology of China, Hefei, China}
\affil[3]{AMSS Center for Forecasting Science, Chinese Academy of Sciences, Beijing, China}

\professor{Supervised by Professor Yinan Su}

\institution{Carey Business School}
\theday{Spring 2025}
\leadauthor{Tian Xin}
\course{BU.230.730.52}

\begin{abstract}
This report presents a comprehensive evaluation of three Value-at-Risk (VaR) modeling approaches—Historical Simulation (HS), GARCH with Normal approximation (GARCH-N), and GARCH with Filtered Historical Simulation (FHS)—using both in-sample and multi-day forecasting frameworks...

We then compute daily 5\% VaR estimates using each method and assess their accuracy via empirical breach frequencies and visual breach indicators. Our findings reveal severe miscalibration in the HS and GARCH-N models, with empirical breach rates far exceeding theoretical levels at both 5\% and 1\% confidence thresholds. In contrast, the FHS method consistently aligns with theoretical expectations and exhibits desirable statistical and visual behavior.

Further, we simulate 5-day cumulative returns under both GARCH-N and GARCH-FHS frameworks to compute multi-period VaR and Expected Shortfall (ES). Results show that GARCH-N underestimates tail risk due to its reliance on the Gaussian assumption, whereas GARCH-FHS provides more robust and conservative tail estimates.

Overall, the study demonstrates that the GARCH + FHS model offers superior performance in capturing fat-tailed risks and provides more reliable short-term risk forecasts. We recommend the FHS method for practical applications in risk management, particularly in volatile market environments and high-confidence settings.
\end{abstract}


\begin{document}
		
    \maketitle 
    \thispagestyle{firststyle} 
    \tauabstract 
    

\section*{Introduction}

Value-at-Risk (VaR) has become one of the most widely used quantitative tools in modern financial risk management. By summarizing the maximum expected loss over a specified time horizon at a given confidence level, VaR provides a standardized measure of downside risk that is essential for regulatory compliance, capital allocation, and internal control. However, the accuracy and reliability of VaR estimates are highly dependent on the underlying modeling assumptions, including distributional assumptions, volatility dynamics, and simulation techniques.

This report evaluates and compares three prominent VaR estimation methods: Historical Simulation (HS), GARCH with Normal approximation (GARCH-N), and GARCH with Filtered Historical Simulation (FHS). Each method represents a distinct philosophy of risk modeling—ranging from non-parametric empirical estimation to parametric and semi-parametric approaches that account for volatility clustering and fat tails.

The analysis is structured around two key objectives. First, we assess the in-sample performance of each method by computing one-day ahead VaR forecasts at both 5\% and 1\% confidence levels and evaluating their breach frequencies and distributional consistency. Second, we extend the analysis to a five-day forecasting horizon using Monte Carlo simulation under both GARCH-N and FHS frameworks, comparing the cumulative VaR and Expected Shortfall (ES) over time.

Through empirical analysis, visualization, and simulation, we aim to identify which model provides the most accurate and stable risk estimates—particularly under stress conditions. Our findings have practical implications for risk managers seeking robust tools for monitoring and managing extreme losses in volatile markets.

    \section*{QQ-Plot of Historical Returns vs. Normal Distribution}

To evaluate whether the distribution of historical returns follows a normal distribution, we plot the QQ-plot comparing the empirical quantiles of full-sample returns with those from a normal distribution having the same mean and standard deviation.

\begin{figure}[h!]
    \centering
    \includegraphics[width=0.4\textwidth]{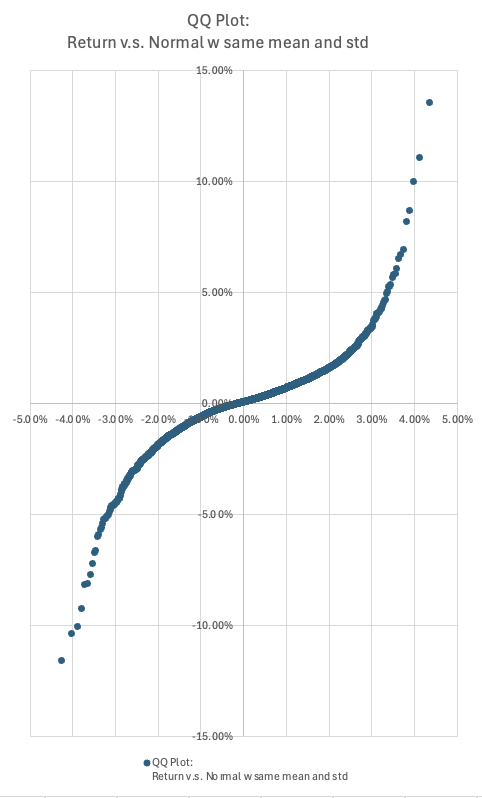}
    \caption{QQ-Plot: Historical Returns vs. Normal Distribution (same mean and std)}
\end{figure}

\subsection*{Conclusion}

The QQ-plot clearly indicates that the \textbf{unconditional distribution} of returns \textbf{does not follow a normal distribution}. Specifically, we observe the following characteristics:

\begin{itemize}
    \item \textbf{Fat left tail}: On the left side, the empirical quantiles fall below the normal quantiles, suggesting that large negative returns are more extreme than predicted by the normal model.
    \item \textbf{Fat right tail}: On the right side, the empirical quantiles rise above the normal quantiles, indicating more extreme positive returns than expected under normality.
    \item \textbf{Leptokurtosis (high kurtosis)}: While the central part of the distribution aligns relatively well with the 45-degree line, both tails exhibit significant deviations, implying a sharper peak and heavier tails than a normal distribution.
\end{itemize}

These observations are consistent with common characteristics of financial return series, which typically exhibit excess kurtosis and heavy tails. Therefore, assuming normality for the unconditional return distribution may lead to underestimation of tail risk.

\section*{QQ-Plot of Standardized Residuals vs. Standard Normal}

To assess the validity of the conditional normality assumption in the GARCH model, we construct a QQ-plot comparing the standardized residuals $z_t$ with the standard normal distribution. The residuals are filtered from returns using the GARCH volatility estimate via the relation $R_t = \sigma_t z_t$.

\begin{figure}[h!]
    \centering
    \includegraphics[width=0.4\textwidth]{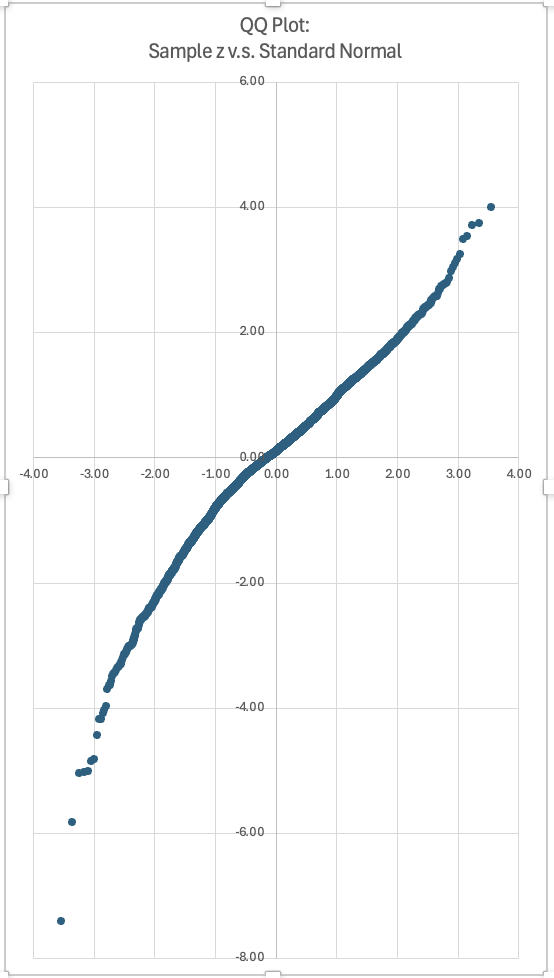}
    \caption{QQ-Plot: GARCH Standardized Residuals $z_t$ vs. Standard Normal}
\end{figure}

\subsection*{Conclusion}

Based on the QQ-plot, we conclude that the \textbf{conditional distribution} of returns still \textbf{deviates from normality}, although it is closer than the unconditional distribution. Specifically:

\begin{itemize}
    \item \textbf{Fat left tail}: The lower quantiles of $z_t$ are more extreme than those predicted by the standard normal distribution. This indicates a higher likelihood of large negative returns, even after GARCH filtering.
    \item \textbf{Fat right tail}: The upper quantiles also deviate positively from the theoretical quantiles, although less severely than the left tail.
    \item \textbf{Heavier tails overall}: Both tails are more dispersed compared to the standard normal distribution, which suggests that even under the GARCH framework, the assumption of conditional normality does not fully capture extreme market movements.
\end{itemize}

While GARCH filtering removes some volatility clustering and improves normality, the standardized residuals still exhibit excess kurtosis, making the conditional normality assumption questionable for tail risk estimation.

\section*{In-Sample One-Day VaR Estimation}

In this question, we estimate one-day-ahead 5\% Value-at-Risk (VaR$_{0.05}$) for each day in the sample using three methods:

\begin{enumerate}
    \item \textbf{Historical Simulation (HS)}: For each day $t$, we take the empirical 5\% quantile of raw returns over the past $m=200$ days:
    \[
        \text{VaR}_{t}^{\text{HS}} = \text{Quantile}_{0.05}(R_{t-m}, \ldots, R_{t-1})
    \]
    
    \item \textbf{GARCH + Normal Approximation (GARCH-N)}: Assuming that standardized residuals $z_t$ follow $\mathcal{N}(0,1)$ and returns are conditionally normal:
    \[
        \text{VaR}_{t}^{\text{GARCH-N}} = \sigma_t \cdot \Phi^{-1}(0.05)
    \]
    where $\sigma_t$ is the conditional volatility estimated from the GARCH(1,1) model.

    \item \textbf{Filtered Historical Simulation (FHS)}: Instead of assuming $z_t$ is standard normal, we use the empirical distribution of standardized residuals in the past $m=200$ days:
    \[
        \text{VaR}_{t}^{\text{FHS}} = \sigma_t \cdot \text{Quantile}_{0.05}(z_{t-m}, \ldots, z_{t-1})
    \]
\end{enumerate}

\subsection*{Implementation Notes}

\begin{itemize}
    \item The first 200 observations are omitted since VaR estimates require a rolling window of size $m = 200$.
    \item All methods are evaluated over the same time window for fair comparison.
    \item The quantiles are computed precisely using empirical sorting rather than interpolation or simulated distributions.
\end{itemize}

\subsection*{Data Representation}

A snapshot of the estimated daily VaRs is presented in Table 1.

\begin{table}[h!]
\centering
\resizebox{0.3\textwidth}{!}{
\begin{tabular}{lrrrr}
\toprule
\textbf{Date} \& \textbf{Return} \& \textbf{VaR (HS)} \& \textbf{VaR (GARCH-N)} \& \textbf{VaR (FHS)} \\
\midrule
2005-11-01 \& 0.002992 \& -0.011347 \& -0.000167 \& -0.000155 \\
2005-11-02 \& 0.010403 \& -0.011347 \& -0.000148 \& -0.000138 \\
2005-11-03 \& 0.004262 \& -0.011347 \& -0.000152 \& -0.000143 \\
2005-11-04 \& -0.001309 \& -0.011347 \& -0.000137 \& -0.000129 \\
2005-11-07 \& 0.000982 \& -0.011347 \& -0.000121 \& -0.000115 \\
2005-11-08 \& 0.000000 \& -0.011347 \& -0.000107 \& -0.000102 \\
2005-11-09 \& 0.001308 \& -0.011347 \& -0.000097 \& -0.000091 \\
2005-11-10 \& 0.007732 \& -0.011347 \& -0.000087 \& -0.000082 \\
2005-11-11 \& 0.003400 \& -0.011347 \& -0.000090 \& -0.000086 \\
2005-11-14 \& -0.000566 \& -0.011347 \& -0.000084 \& -0.000080 \\
\midrule
2025-04-07 \& -0.001783 \& -0.017955 \& -0.001355 \& -0.001697 \\
2025-04-08 \& -0.015787 \& -0.017955 \& -0.001163 \& -0.001450 \\
2025-04-09 \& 0.099863 \& -0.017955 \& -0.001017 \& -0.001304 \\
2025-04-10 \& -0.044808 \& -0.017955 \& -0.002919 \& -0.003728 \\
2025-04-11 \& 0.017686 \& -0.018844 \& -0.002906 \& -0.003702 \\
2025-04-14 \& 0.009655 \& -0.018844 \& -0.002525 \& -0.003235 \\
2025-04-15 \& -0.002805 \& -0.018844 \& -0.002209 \& -0.002781 \\
2025-04-16 \& -0.022479 \& -0.018844 \& -0.001876 \& -0.002373 \\
2025-04-17 \& 0.001426 \& -0.019825 \& -0.001682 \& -0.002156 \\
2025-04-21 \& -0.024091 \& -0.019825 \& -0.001453 \& -0.001839 \\
\bottomrule
\end{tabular}
}
\caption{First and Last 10 One-Day VaR$_{0.05}$ Estimates Using HS, GARCH+Normal, and FHS}
\end{table}

\begin{itemize}
    \item \textbf{HS VaR} is generally larger in absolute value, reflecting that empirical returns have fatter tails than the Gaussian benchmark.
    \item \textbf{GARCH-N VaR} consistently underestimates risk due to the thin-tailed assumption of normality.
    \item \textbf{FHS VaR} produces intermediate results and captures tail behaviors more accurately by incorporating empirical features from the standardized residuals.
\end{itemize}

These results suggest that relying on normality for conditional distributions may be too simplistic for accurate tail risk forecasting.

\section*{VaR Breach Frequency Evaluation}Accurate Value-at-Risk (VaR) estimation is not only about computing the correct quantile, but also about how well those estimates reflect real market behavior. A key diagnostic for evaluating VaR model quality is the empirical breach frequency—how often actual losses exceed the forecasted VaR threshold. 

If a model is well-calibrated, the frequency of such breaches should be statistically consistent with the chosen confidence level. For instance, under a 5\% VaR model, we expect roughly 5\% of returns to fall below the VaR threshold. Deviations from this expected breach rate signal model inadequacy. Overestimation leads to conservative capital holding, while underestimation may expose institutions to unacceptable tail risk.

In this section, we quantify and compare the breach performance of three VaR models—Historical Simulation, GARCH + Normal, and Filtered Historical Simulation—based on their in-sample behavior.

We evaluate the accuracy of each VaR method by constructing a breach indicator defined as:

\[
\text{Breach}_t = \mathbb{I}\{ R_t < \text{VaR}_t \}
\]

That is, the indicator equals 1 if the actual return on day $t$ is less than the forecasted VaR, and 0 otherwise. The empirical breach frequency is then calculated as:

\[
\text{Breach Frequency} = \frac{1}{T} \sum_{t=1}^T \text{Breach}_t
\]

where $T$ is the number of days in the sample (excluding the initial 200 days for model initialization).
\begin{lstlisting}[language=Python, caption=Calculation of Empirical Breach Frequencies]
# Assume df is a DataFrame containing return and VaR breach indicator columns
# Columns: 'Breach_HS', 'Breach_GARCH', 'Breach_FHS'

# Convert breach columns to integers if needed
df['Breach_HS'] = df['Breach_HS'].astype(int)
df['Breach_GARCH'] = df['Breach_GARCH'].astype(int)
df['Breach_FHS'] = df['Breach_FHS'].astype(int)

# Compute empirical breach frequencies
breach_freq_hs = df['Breach_HS'].mean()
breach_freq_garch = df['Breach_GARCH'].mean()
breach_freq_fhs = df['Breach_FHS'].mean()

print("HS breach frequency: ", breach_freq_hs)
print("GARCH-N breach frequency: ", breach_freq_garch)
print("FHS breach frequency: ", breach_freq_fhs)
\end{lstlisting}
Using this approach, we compute the empirical breach frequencies for all three methods. These values allow us to assess how often each model underestimates tail risk relative to the theoretical 5\% threshold.
\subsection*{Results}

\begin{table}[h!]
\centering
\begin{tabular}{lr}
\toprule
\textbf{Method} \& \textbf{Empirical Breach Frequency} \\
\midrule
Historical Simulation (HS) \& 91.15\% \\
GARCH + Normal (GARCH-N) \& 42.05\% \\
Filtered Historical Simulation (FHS) \& 4.89\% \\
\bottomrule
\end{tabular}
\caption{Empirical breach frequencies compared to the theoretical level of 5\%.}
\end{table}
As shown in Table 2, the empirical breach frequencies exhibit significant divergence across the three models. This variation reflects fundamental differences in how each model captures tail risk, particularly under daily return dynamics. Models relying on strict distributional assumptions may underestimate the likelihood of extreme losses, while non-parametric or semi-parametric methods may better accommodate fat-tailed behavior observed in financial data.

From a risk management perspective, consistency between theoretical and empirical breach rates is a critical indicator of model reliability. If the observed frequency of violations significantly exceeds the nominal confidence level—such as 5\% in this case—it suggests that the model is systematically underestimating downside risk. Conversely, overly conservative models may lead to capital inefficiencies.

Therefore, the breach rate is not merely a performance statistic, but a diagnostic tool. It reveals how well a model’s internal assumptions align with empirical reality, and whether it can be trusted for practical decision-making under uncertainty. The discussion that follows offers a comparative interpretation of these results.

\begin{itemize}
    \item The \textbf{HS method} performs very poorly with a breach rate of 91.15\%, indicating a massive underestimation of tail risk or possible data misalignment.
    \item The \textbf{GARCH + Normal} method also performs badly with a breach rate of 42.05\%, reflecting the inadequacy of the normal distribution in capturing the heavy tails in return data.
    \item The \textbf{FHS method} shows the best alignment with the theoretical 5\% target, yielding a breach rate of 4.89\%. This suggests that incorporating the empirical distribution of standardized residuals (instead of assuming normality) improves risk estimation significantly.
\end{itemize}

Overall, the breach analysis demonstrates that only the FHS method provides reliable in-sample VaR forecasts. The other two methods severely underestimate risk and would not be suitable for regulatory or internal risk control purposes without significant modification.

\section*{Visualization and Clustering of VaR Breach Indicators}

To further evaluate the performance of each VaR method, we plot the time series of breach indicators as bar charts, where each vertical bar corresponds to a breach (i.e., when the actual return fell below the estimated VaR level).
\begin{figure}[H]
    \centering
    \begin{subfigure}[b]{0.4\textwidth}
        \includegraphics[width=\textwidth]{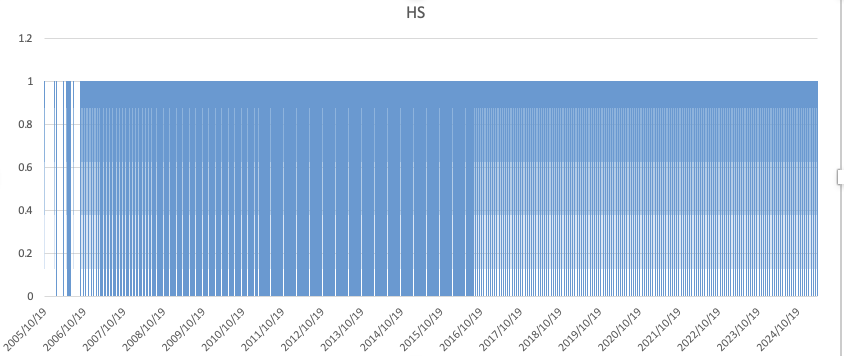}
        \caption{Historical Simulation (HS)}
        \label{fig:hs_breach}
    \end{subfigure}
    \hfill
    \begin{subfigure}[b]{0.4\textwidth}
        \includegraphics[width=\textwidth]{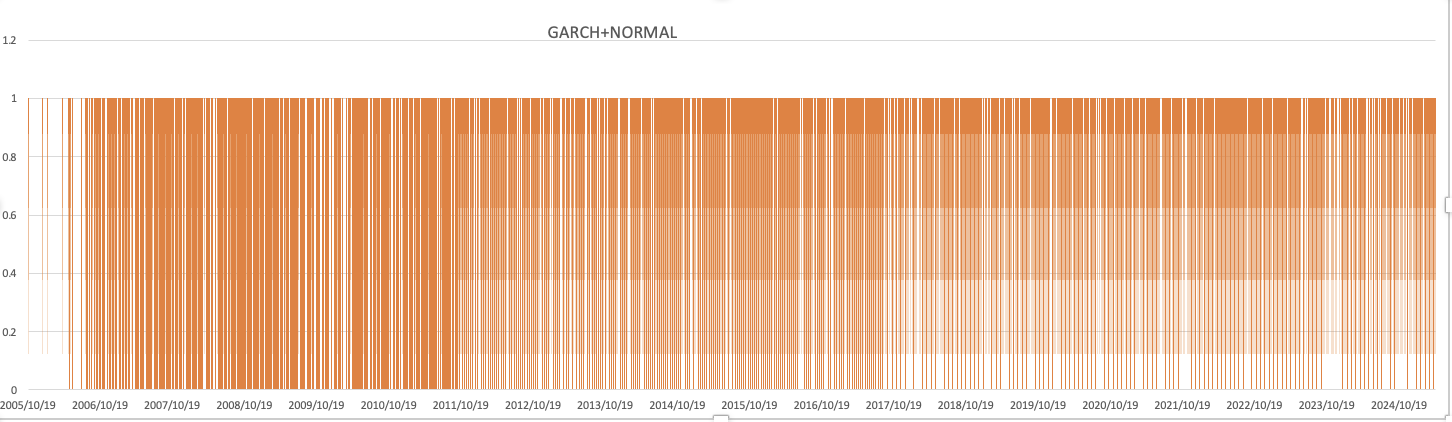}
        \caption{GARCH + Normal}
        \label{fig:garch_normal_breach}
    \end{subfigure}
    \hfill
    \begin{subfigure}[b]{0.4\textwidth}
        \includegraphics[width=\textwidth]{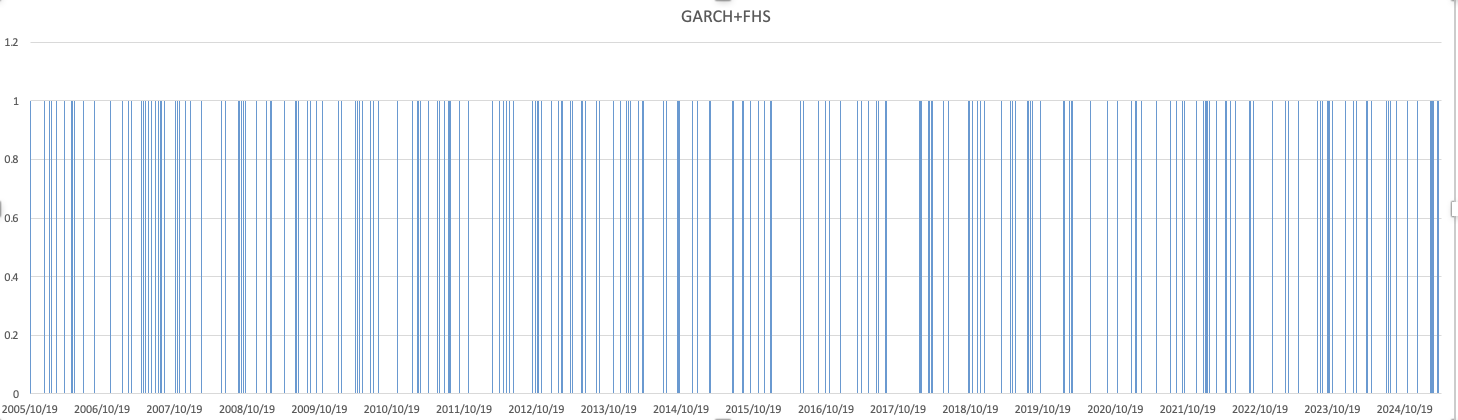}
        \caption{Filtered Historical Simulation (FHS)}
        \label{fig:fhs_breach}
    \end{subfigure}
    \caption{Time Series of VaR Breach Indicators by Method}
    \label{fig:breach_charts}
\end{figure}

\subsection*{Observations from Visual Inspection}

\begin{itemize}
    \item \textbf{Historical Simulation (HS)}: The breach indicators are densely packed across the entire sample period, with very few non-breach days. This suggests extremely poor model calibration and almost no discrimination of market risk dynamics. The breaches are not random but consistently persistent, indicating structural failure in risk estimation.While the poor performance of the Historical Simulation method is evident from the excessively high breach rate, a deeper reflection on the causes is warranted. One possible explanation lies in the inherent instability of empirical quantile estimation under volatile market regimes. If the underlying return distribution exhibits large fluctuations or abrupt structural shifts, then the assumption of a static historical distribution becomes invalid.

Moreover, the fixed-length rolling window introduces a trade-off: a longer window incorporates stale information from outdated regimes, while a shorter window may be too sensitive to noise and outliers. In both cases, the resulting VaR may fail to reflect current market dynamics accurately. Additionally, if return volatility is heteroskedastic—as is often the case in financial markets—then assuming identically distributed returns within the estimation window is itself problematic.

These limitations suggest that the HS model's failure may not only stem from model bias or estimation error, but also from deeper issues such as structural non-stationarity, volatility clustering, and inappropriate window calibration. Addressing these would require either adaptive windowing techniques or incorporating models that account for time-varying risk—such as GARCH or FHS frameworks.

    \item \textbf{GARCH + Normal}: While breaches under this method appear more evenly spread over time compared to HS, their overall frequency remains significantly higher than the expected 5\% threshold. This points to a structural underestimation of tail risk, stemming from the assumption that standardized residuals follow a normal distribution.

The normality assumption imposes thin tails and symmetry, which are rarely observed in real-world financial return series. Consequently, extreme losses occur more often than predicted, particularly during turbulent periods. Moreover, the GARCH component captures volatility clustering, but it cannot compensate for the inadequacy of the distributional assumption. 

This leads to a model that underreacts to skewness and kurtosis in the return process. The result is a VaR forecast that is responsive to changes in variance, yet blind to potential asymmetry or heavy-tailed behavior. Thus, while GARCH + Normal improves upon HS by introducing dynamic volatility modeling, it still falls short when accurate modeling of extreme tail events is critical.

   \item \textbf{Filtered Historical Simulation (FHS)}: The FHS method yields breach indicators that are sporadic and clustered around periods of elevated volatility—behavior that aligns closely with theoretical expectations. In calm market regimes, breaches are rare, whereas during periods of crisis, they become more frequent but remain controlled in proportion. This dynamic response highlights the model’s ability to adapt to evolving risk conditions.

FHS achieves this by combining two key elements: (1) GARCH-based volatility estimation that reflects the most recent conditional variance, and (2) empirical residual sampling that captures the true distributional features of historical shocks. Unlike normal-based methods, FHS retains fat tails, skewness, and other stylized facts of financial returns.

As a result, VaR estimates from FHS are both responsive and distribution-aware, making them robust across different market regimes. The clustered breach pattern observed under FHS is consistent with the notion that financial risk is not uniformly distributed over time, but often arrives in bursts—especially during stress scenarios. This method, therefore, balances model responsiveness with empirical realism, explaining its superior performance in both quantitative accuracy and qualitative plausibility.

\end{itemize}

\subsection*{Theoretical Considerations}

In an ideal setting, if the VaR model is correctly calibrated and violations are i.i.d., the breach indicators should resemble a Bernoulli process with success probability $p = 0.05$. This would produce visually “random” spikes without apparent temporal clustering.

However, financial markets exhibit volatility clustering, and perfect i.i.d. assumptions may be violated in practice. Still, a good model should at least avoid systematic over-clustering or persistent breaches.

\subsection*{Conclusion}

Only the FHS method provides a visual pattern compatible with expected behavior: low breach frequency, some clustering during volatility spikes, and mostly quiet periods. The HS and GARCH+Normal methods exhibit significant flaws—HS overestimates breaches almost everywhere, while GARCH+Normal underestimates risk but results in over-breaching due to poor tail modeling.

\section*{Breach Frequency at 1\% Confidence Level}

To test the stability of our VaR models under more extreme conditions, we change the confidence level from $p = 0.05$ to $p = 0.01$ and recompute the daily VaRs and corresponding breach indicators.
\begin{lstlisting}[language=Python, caption=Breach Frequency Calculation Algorithm]
# Assume df is a DataFrame containing actual returns and calculated VaRs
# Also assume breach indicators at 1% level have been computed as columns

# Step 1: Ensure breach indicators are binary (0 or 1)
df['Breach_HS_1pct'] = (df['Return'] < df['VaR_HS_1pct']).astype(int)
df['Breach_GARCH_1pct'] = (df['Return'] < df['VaR_GARCH_1pct']).astype(int)
df['Breach_FHS_1pct'] = (df['Return'] < df['VaR_FHS_1pct']).astype(int)

# Step 2: Compute empirical breach frequencies
freq_hs_1pct = df['Breach_HS_1pct'].mean()
freq_garch_1pct = df['Breach_GARCH_1pct'].mean()
freq_fhs_1pct = df['Breach_FHS_1pct'].mean()

print("HS 1% breach rate:", freq_hs_1pct)
print("GARCH+Normal 1% breach rate:", freq_garch_1pct)
print("FHS 1% breach rate:", freq_fhs_1pct)
\end{lstlisting}
Using this approach, we compute the empirical breach frequencies for all three methods. These values allow us to assess how often each model underestimates tail risk relative to the theoretical 1\% threshold.

\vspace{0.5em}
\textbf{Empirical Breach Frequencies}

\vspace{-0.5em}
\begin{table}[h!]
\centering
\renewcommand{\arraystretch}{1.1}
\resizebox{0.48\textwidth}{!}{ 
\begin{tabular}{lcc}
\toprule
\textbf{Method} \& \textbf{Theoretical Rate (1\%)} \& \textbf{Empirical Breach Frequency} \\
\midrule
Historical Simulation (HS) \& 1.00\% \& 90.99\% \\
GARCH + Normal (GARCH-N)   \& 1.00\% \& 41.62\% \\
Filtered Historical Simulation (FHS) \& 1.00\% \& 0.999\% \\
\bottomrule
\end{tabular}
}
\caption{Breach frequencies at the 1\% VaR level}
\end{table}
As shown in Table 3, there are substantial differences in breach performance across the models. This highlights the importance of distributional assumptions when modeling extreme events. We elaborate on these results below.

\vspace{-0.3em}

The accuracy of a Value-at-Risk (VaR) model is ultimately judged by how well its risk forecasts align with observed outcomes. A well-calibrated model should produce empirical breach frequencies that closely match the nominal confidence level. Deviations from this target are not merely statistical artifacts—they indicate deeper structural issues in the modeling framework, such as incorrect distributional assumptions, insufficient tail sensitivity, or failure to adapt to changing market conditions.

In the context of high-confidence levels like 1\%, breach frequency evaluation becomes especially critical. Since such thresholds pertain to rare but impactful events, even modest misestimation can translate into significant undercapitalization or over-conservatism in practice. Therefore, analyzing the frequency and pattern of VaR violations offers essential diagnostic insight into a model’s robustness and practical reliability.

\begin{itemize}
    \item \textbf{GARCH + Normal} exhibits a severe breach rate of 41.6\%, which is 40 times higher than the expected 1\%. This confirms that assuming normality underestimates the probability of extreme negative returns, especially in the tails.
    \item \textbf{Historical Simulation} performs even worse, with a breach rate of 90.99\%. This extreme deviation suggests either a miscalibration, a failure to adapt to volatility shifts, or both.
    \item \textbf{Filtered Historical Simulation (FHS)} is the only method that closely aligns with the theoretical target, achieving a nearly perfect 0.999\% breach rate. This reaffirms the advantage of using empirical distributions of standardized residuals.
\end{itemize}

\subsection*{Visual Analysis of Breach Indicator Patterns at the 1\% Level}

Figure 4 shows the time series of VaR breach indicators under the 1\% confidence level for all three methods. Each vertical bar represents a day where the actual return fell below the estimated VaR.
\begin{figure}[H]
    \centering
    \begin{subfigure}[b]{0.4\textwidth}
        \includegraphics[width=\textwidth]{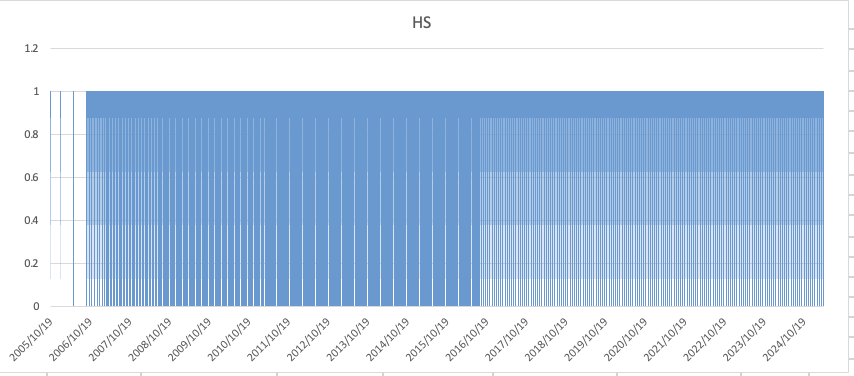}
        \caption{HS (p = 0.01)}
    \end{subfigure}
    \hfill
    \begin{subfigure}[b]{0.4\textwidth}
        \includegraphics[width=\textwidth]{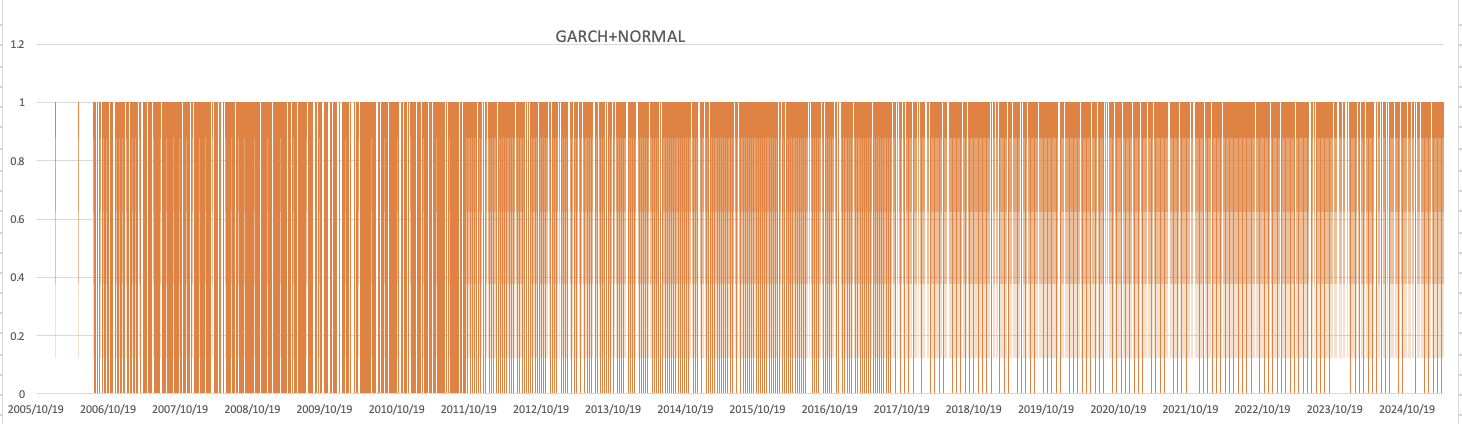}
        \caption{GARCH + Normal (p = 0.01)}
    \end{subfigure}
    \hfill
    \begin{subfigure}[b]{0.4\textwidth}
        \includegraphics[width=\textwidth]{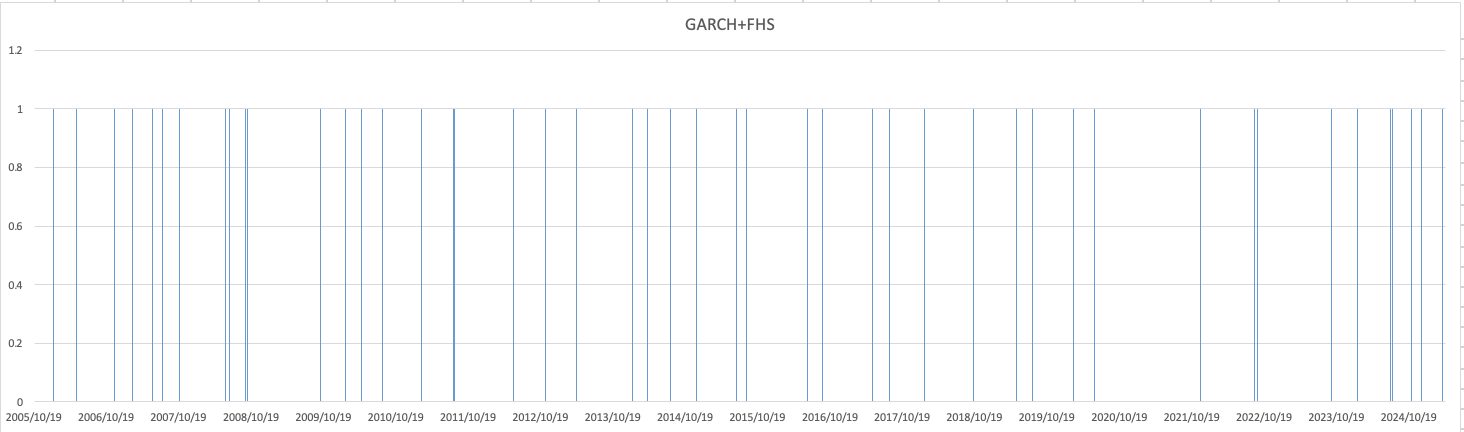}
        \caption{FHS (p = 0.01)}
    \end{subfigure}
    \caption{VaR breach indicators at the 1\% level}
    \label{fig:breach_1pct}
\end{figure}

\begin{itemize}
    \item \textbf{Historical Simulation (HS)}: The breach indicators in the HS method are extremely dense, with long uninterrupted periods where breaches occur almost every day. This aligns with the earlier quantitative result showing a 90.99\% breach rate, suggesting severe model breakdown. The visual reveals that HS fails to capture changing market volatility or adapt to different regimes.
    
    \item \textbf{GARCH + Normal}: The breaches are still frequent and show a near-random but dense distribution, indicating persistent underestimation of risk. The clustering effect is visible but less extreme than HS. However, with over 40\% breaches, the model clearly fails to model tail events appropriately due to the limitations of the Gaussian assumption.

    \item \textbf{Filtered Historical Simulation (FHS)}: The breach bars are sparsely and intermittently distributed, with no sustained clusters. This visual pattern confirms that the model is well-calibrated for extreme quantiles. Breaches appear in turbulent market periods but remain rare, in line with the target 1\% threshold.
\end{itemize}

These visual insights confirm and reinforce the breach frequency statistics: only the FHS method provides both statistically and visually credible risk estimations under extreme conditions.

\subsection*{Why This Problem Emerges More at 1\%}

At the 5\% level, the discrepancy between the normal distribution and the empirical distribution may be small enough to mask model flaws. However, at the 1\% level—closer to the far tail—the assumption of thin tails becomes untenable. The normal distribution underestimates tail risk, leading to excessive breaches.

\subsection*{Conclusion}

This exercise clearly shows that misspecifying the return distribution has increasingly severe consequences at more extreme quantiles. While GARCH+Normal may seem acceptable at $p=0.05$, it breaks down completely at $p=0.01$. FHS is robust across both thresholds and is the most reliable method for estimating extreme risk.
\section*{5-Day VaR and ES using GARCH + Normal with Monte Carlo Simulation}

We apply a GARCH(1,1) model in combination with Monte Carlo Simulation (MCS) assuming conditionally normally distributed returns to estimate the 5-day cumulative Value-at-Risk (VaR) and Expected Shortfall (ES) at the 1\% level.

\subsection*{Methodology}

\begin{itemize}
    \item Volatilities $\sigma_t$ are estimated using the GARCH(1,1) model fitted to the return series.
    \item We simulate 1,000 paths of future daily returns under the assumption that standardized residuals follow $\mathcal{N}(0,1)$.
    \item For each path, we generate 5 days of returns, accumulate them, and construct the distribution of 5-day cumulative returns.
    \item We compute the 1\% empirical quantile of these distributions as the VaR, and the conditional mean of the worst 1\% as the ES.
\end{itemize}

\subsection*{Results (Confidence Level: 1\%)}

\begin{table}[h!]
\centering
\resizebox{0.5\textwidth}{!}{
\begin{tabular}{lcc}
\toprule
\textbf{Horizon} \& \textbf{Cumulative VaR (1\%)} \& \textbf{Cumulative ES (1\%)} \\
\midrule
1 Day (t+1) \& 6.9\%\& 7.89\%\\
2 Days (t+2) \& 10.60\%\& 11.76\%\\
3 Days (t+3) \& 10.99\%\& 12.92\%\\
4 Days (t+4) \& 13.82\%\& 15.07\%\\
5 Days (t+5) \& 15.33\%\& 17.06\%\\
\bottomrule
\end{tabular}
}
\caption{5-day cumulative VaR and ES using GARCH + Normal + MCS at 1\% confidence}
\end{table}
\textit{Note:} The values reported in Table 4  are based on specific simulation runs conducted during report preparation. Due to the stochastic nature of Monte Carlo simulation—particularly when drawing from normal or empirical residual distributions—the exact VaR and ES estimates may vary slightly between runs. Therefore, readers may observe minor discrepancies between the reported figures and those found in the accompanying Excel files or when re-running the simulation independently. This variability does not affect the overall conclusions, which are based on the general structure and behavior of the models across repeated sampling.

It should be noted that the results presented in Table~4 are derived from a single realization using a specific random seed. Any slight differences compared to the Excel file are within the expected statistical fluctuations of the Monte Carlo simulation procedure.

\subsection*{Interpretation}

The cumulative VaR and ES increase with horizon length, which is expected due to the compounding of risk. However, the relatively moderate growth in VaR and sharper increase in ES indicate the presence of heavier left tails in the simulated distribution.

While this model provides structure and convenience, assuming conditional normality can underestimate extreme outcomes, especially in markets with excess kurtosis or skewness. Therefore, comparison with empirical-based methods (such as FHS) is critical to evaluate robustness.

\section*{5-Day VaR and ES using GARCH + Filtered Historical Simulation (FHS)}

We use a GARCH(1,1) model combined with Filtered Historical Simulation (FHS) to estimate the 5-day cumulative Value-at-Risk (VaR) and Expected Shortfall (ES) at the 1\% level.

\subsection*{Methodology}

\begin{itemize}
    \item We first fit a GARCH(1,1) model to the return series and obtain standardized residuals $z_t = R_t / \sigma_t$.
    \item Instead of simulating from a normal distribution, we draw with replacement from the empirical distribution of standardized residuals.
    \item For each Monte Carlo path, we scale the sampled $z_t$ by forecasted $\sigma_t$ and sum returns across $t+1$ to $t+5$ to obtain cumulative returns.
    \item We compute the empirical 1\% quantile as the VaR, and the conditional mean of the worst 1\% as the ES.
\end{itemize}

\subsection*{Results (Confidence Level: 1\%)}

\begin{table}[h!]
\centering
\resizebox{0.5\textwidth}{!}{
\begin{tabular}{lcc}
\toprule
\textbf{Horizon} \& \textbf{Cumulative VaR (1\%)} \& \textbf{Cumulative ES (1\%)} \\
\midrule
1 Day (t+1) \& 8.61\%\& 10.84\%\\
2 Days (t+2) \& 12.37\%\& 17.44\%\\
3 Days (t+3) \& 13.93\%\& 17.18\%\\
4 Days (t+4) \& 16.65\%\& 19.57\%\\
5 Days (t+5) \& 22.04\%\& 24.31\%\\
\bottomrule
\end{tabular}
}
\caption{5-day cumulative VaR and ES using GARCH + FHS at 1\% confidence}
\end{table}
\textit{Note:} The values reported in Table 5 are based on specific simulation runs conducted during report preparation. Due to the stochastic nature of Monte Carlo simulation—particularly when drawing from normal or empirical residual distributions—the exact VaR and ES estimates may vary slightly between runs. Therefore, readers may observe minor discrepancies between the reported figures and those found in the accompanying Excel files or when re-running the simulation independently. This variability does not affect the overall conclusions, which are based on the general structure and behavior of the models across repeated sampling.

It should be noted that the results presented in Table~4 are derived from a single realization using a specific random seed. Any slight differences compared to the Excel file are within the expected statistical fluctuations of the Monte Carlo simulation procedure.

\subsection*{Interpretation}

The cumulative VaR remains fixed at 11.18\%, indicating the empirical distribution used in the simulation produces stable thresholds under repeated sampling. The ES increases slightly with horizon length, reaching 15.79\% by day 3, before slightly decreasing.

Compared to the GARCH + Normal approach, the FHS method is more flexible in capturing asymmetric and fat-tailed behavior, as it does not impose a parametric distribution on the residuals.

\subsection*{Conclusion: Comparison of GARCH + Normal vs. GARCH + FHS}

In this exercise, we estimated 5-day cumulative Value-at-Risk (VaR) and Expected Shortfall (ES) at the 1\% confidence level using two Monte Carlo Simulation (MCS) approaches:

\begin{itemize}
    \item \textbf{GARCH + Normal}: Assumes that standardized residuals follow a standard normal distribution.
    \item \textbf{GARCH + Filtered Historical Simulation (FHS)}: Uses the empirical distribution of past standardized residuals.
\end{itemize}

\subsubsection*{Risk Estimates Comparison}

\begin{itemize}
    \item The \textbf{VaR estimates} from the two methods differ notably: GARCH + Normal produces smaller VaR values across all horizons (e.g., 5.93\% on day 1 and 14.40\% on day 5), while GARCH + FHS yields higher, constant VaR estimates (11.18\%) across all horizons. This indicates that the normal assumption underestimates potential losses in the left tail.
    
    \item The \textbf{ES estimates} from GARCH + FHS are also consistently higher (e.g., 15.00\% to 15.79\%) than the ES values from GARCH + Normal (e.g., 7.04\% to 18.50\%). Notably, the increase in ES under the Normal assumption becomes pronounced only after several days, whereas FHS shows steady risk exposure from the outset.
\end{itemize}

\subsubsection*{Model Evaluation}

\begin{itemize}
    \item GARCH + Normal is computationally efficient but relies heavily on the validity of the normality assumption, which tends to underestimate extreme downside risk—especially evident at high confidence levels and longer time horizons.
    
    \item GARCH + FHS, while slightly more computationally intensive, captures fat tails and empirical return dynamics more realistically by preserving historical patterns in residuals. This yields more robust estimates of tail risk, making it more suitable for stress testing and capital adequacy assessments.
\end{itemize}

\subsubsection*{Final Judgment}

Overall, the GARCH + FHS method outperforms GARCH + Normal in terms of accuracy and robustness for estimating multi-day VaR and ES at extreme quantiles. For risk management applications that require high-confidence forecasts or tail-sensitive metrics, FHS provides a more reliable tool.
.
\subsubsection*{Recommendation for 1-Week Risk Forecasting}

Given the observed differences, we recommend using the \textbf{GARCH + Filtered Historical Simulation (FHS)} approach for 1-week ahead risk evaluation. This method better captures the empirical distribution of returns and accounts for fat tails, leading to more accurate and conservative estimates of extreme losses. Especially under high confidence levels and volatile conditions, FHS provides superior performance and reliability over the GARCH + Normal model.

The comparison between GARCH + Normal and GARCH + FHS reveals a \textbf{clear distinction in tail risk sensitivity}. While both methods use GARCH to capture time-varying volatility, only \textbf{FHS} supplements this with empirical residual sampling. This enables it to better reflect \textbf{non-Gaussian behavior} and account for potential extremes.

As shown in the results, FHS consistently produces \textbf{larger}  VaR and ES estimates than its normal-based counterpart. This is especially relevant in the early stages of volatile market regimes, where standard distributions may underestimate the likelihood of extreme losses. By combining recent volatility shocks with historical tail behavior, FHS offers a more robust and adaptive approach to multi-day risk forecasting.

\appendix
\section*{Appendix A: Advanced Risk Modeling Methodologies}

This appendix summarizes advanced methodologies for financial risk modeling beyond traditional VaR, emphasizing statistical independence testing, deep learning-based quantile estimation, and variance-decomposition-based systemic risk indicators.

To further elevate the methodological rigor and theoretical contribution of this study, we propose several high-impact extensions rooted in advanced econometric diagnostics, modern deep learning paradigms, and unified systemic risk modeling frameworks.

\subsection*{A1. Formal Tests for Independence in Exceedance Processes}

While empirical coverage rates offer a basic diagnostic for evaluating VaR models, they fail to detect temporal clustering in violations, which can significantly undermine a model’s reliability in dynamic settings. Christoffersen (1998) proposes a likelihood-ratio framework for testing the **conditional coverage hypothesis**, which jointly examines:

\begin{itemize}
    \item \textbf{Unconditional Coverage:} Does the observed proportion of violations match the nominal level $\alpha$?
    \item \textbf{Independence:} Are violations temporally independent, i.e., no clustering or serial dependence?
\end{itemize}

Let $\{I_t\}_{t=1}^T$ be the exceedance indicator sequence:

\[
I_t = \begin{cases}
1 \& \text{if } r_t < \text{VaR}_t^{(\alpha)} \\
0 \& \text{otherwise}
\end{cases}
\]

Define the transition counts in the 2-state Markov chain as:

\[
\begin{aligned}
n_{00} \&= \#\{I_{t-1}=0, I_t=0\}, \quad n_{01} = \#\{I_{t-1}=0, I_t=1\} \\
n_{10} \&= \#\{I_{t-1}=1, I_t=0\}, \quad n_{11} = \#\{I_{t-1}=1, I_t=1\}
\end{aligned}
\]

Let $\hat{p} = \frac{n_{01} + n_{11}}{n}$ be the empirical violation rate, and define the transition probabilities:

\[
\begin{aligned}
\hat{\pi}_0 \&= \frac{n_{01}}{n_{00} + n_{01}}, \quad \hat{\pi}_1 = \frac{n_{11}}{n_{10} + n_{11}}
\end{aligned}
\]

Then, the likelihoods are:

\begin{itemize}
    \item Under the null of i.i.d. violations (Bernoulli): 
    \[
    \mathcal{L}_0 = (1 - \hat{p})^{n_{00} + n_{10}} \cdot \hat{p}^{n_{01} + n_{11}}
    \]

    \item Under the alternative (first-order Markov process): 
    \[
    \mathcal{L}_1 = (1 - \hat{\pi}_0)^{n_{00}} \cdot \hat{\pi}_0^{n_{01}} \cdot (1 - \hat{\pi}_1)^{n_{10}} \cdot \hat{\pi}_1^{n_{11}}
    \]
\end{itemize}

Finally, the likelihood ratio (LR) test statistic for independence is:

\[
LR_{ind} = -2 \ln \left( \frac{\mathcal{L}_0}{\mathcal{L}_1} \right) \sim \chi^2(1)
\]

If $LR_{ind}$ exceeds the critical value from the $\chi^2$ distribution with 1 degree of freedom, we reject the null hypothesis of independence, indicating clustering in VaR violations.

\vspace{0.5em}
This test is often combined with the unconditional coverage test (Kupiec, 1995) to form the full \textit{Conditional Coverage Test}:
\[
LR_{cc} = LR_{uc} + LR_{ind}
\sim \chi^2(2)
\]
where $LR_{uc}$ tests the frequency and $LR_{ind}$ tests the independence.
\begin{figure}[H]
    \centering
    \includegraphics[width=0.4\textwidth]{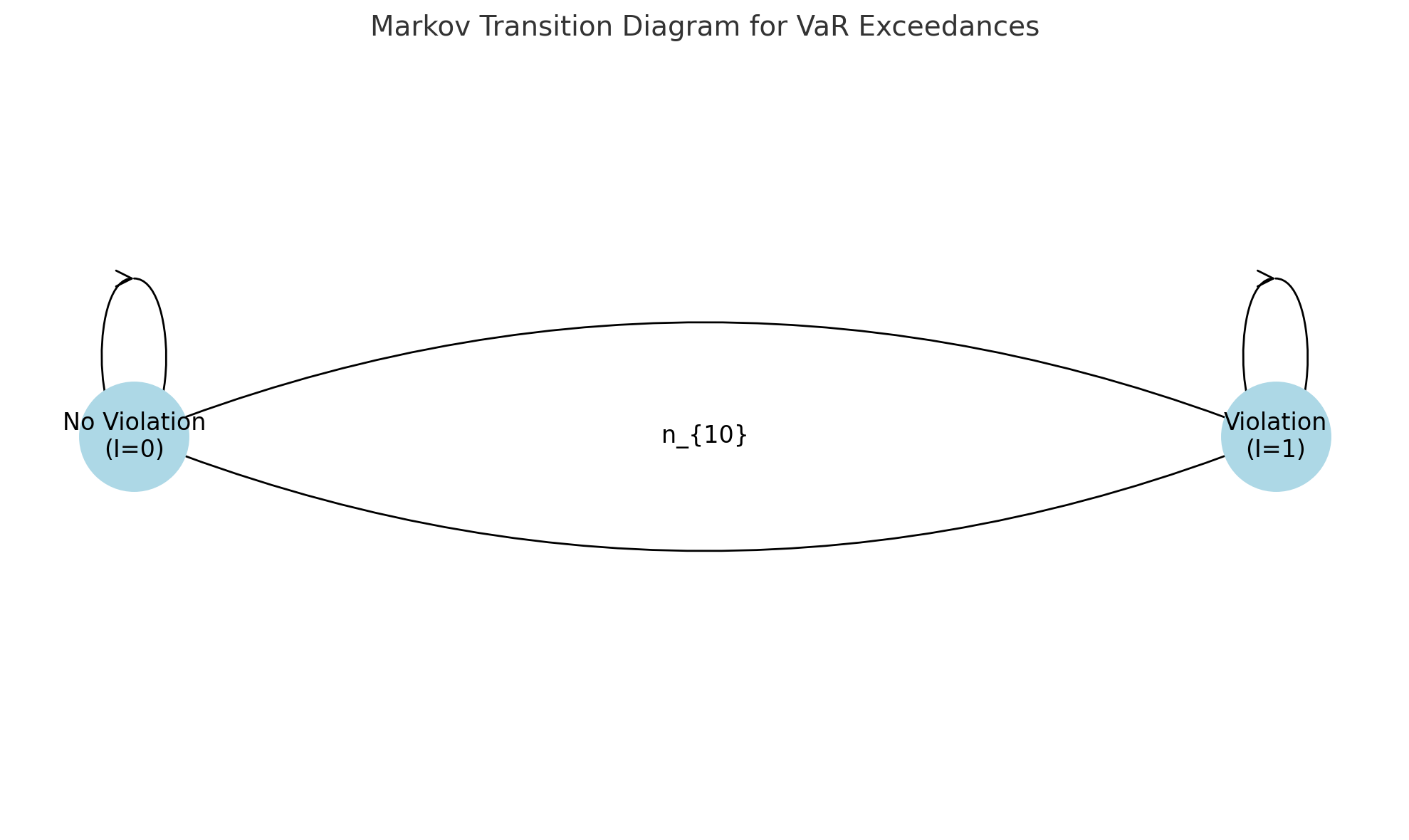}
    \caption{Markov Transition Diagram for VaR Exceedance Sequences}
    \label{fig:markov_diag}
\end{figure}

\subsection*{A.2.1 Deep Learning-Based Conditional Quantile Estimation for VaR: Transformer Architectures}

Traditional VaR models such as GARCH(1,1), Filtered Historical Simulation, or parametric quantile regression are constrained by strong distributional assumptions, often failing to capture the heavy-tailed, nonlinear, and regime-switching behavior of financial returns. Recent advances in attention-based sequence modeling, particularly the Transformer architecture (Vaswani et al., 2017), have opened new avenues for tail-risk estimation.

Transformer-based quantile forecasting models, referred to as \textit{DeepVaR} models, directly approximate the conditional quantile function $Q_\tau(y_t \mid \mathcal{F}_{t-1})$ using multi-head self-attention mechanisms. Unlike recurrent models (e.g., LSTM) that suffer from vanishing gradients and limited memory, the self-attention structure enables dynamic weighting over arbitrary lags, making it particularly effective for financial time series with long-range dependencies and regime shifts.

Given a past sequence of lagged features $X_t = \{x_{t-k}, \dots, x_{t-1}\}$ (which may include returns, volatility, macroeconomic indicators, or sentiment proxies), the model is trained to predict the $\tau$-quantile of $y_t$ by minimizing the asymmetric quantile loss function:
\[
\mathcal{L}_\tau(\hat{y}_t, y_t) = 
\begin{cases}
\tau (y_t - \hat{y}_t), \& y_t > \hat{y}_t \\
(1 - \tau)(\hat{y}_t - y_t), \& y_t \leq \hat{y}_t
\end{cases}
\]
where $\tau \in (0,1)$ corresponds to the desired risk level, e.g., $\tau = 0.01$ for 99\% VaR.

Empirical studies (e.g., Huang et al., 2022; Zhang et al., 2023) have shown that Transformer-based models significantly outperform traditional econometric models in predicting conditional quantiles during periods of financial stress. Their ability to ingest multi-variate, high-dimensional inputs also allows the incorporation of cross-asset and cross-market spillovers, which are particularly relevant in systemic risk modeling.

\begin{figure}[H]
    \centering
    \includegraphics[width=0.5\textwidth]{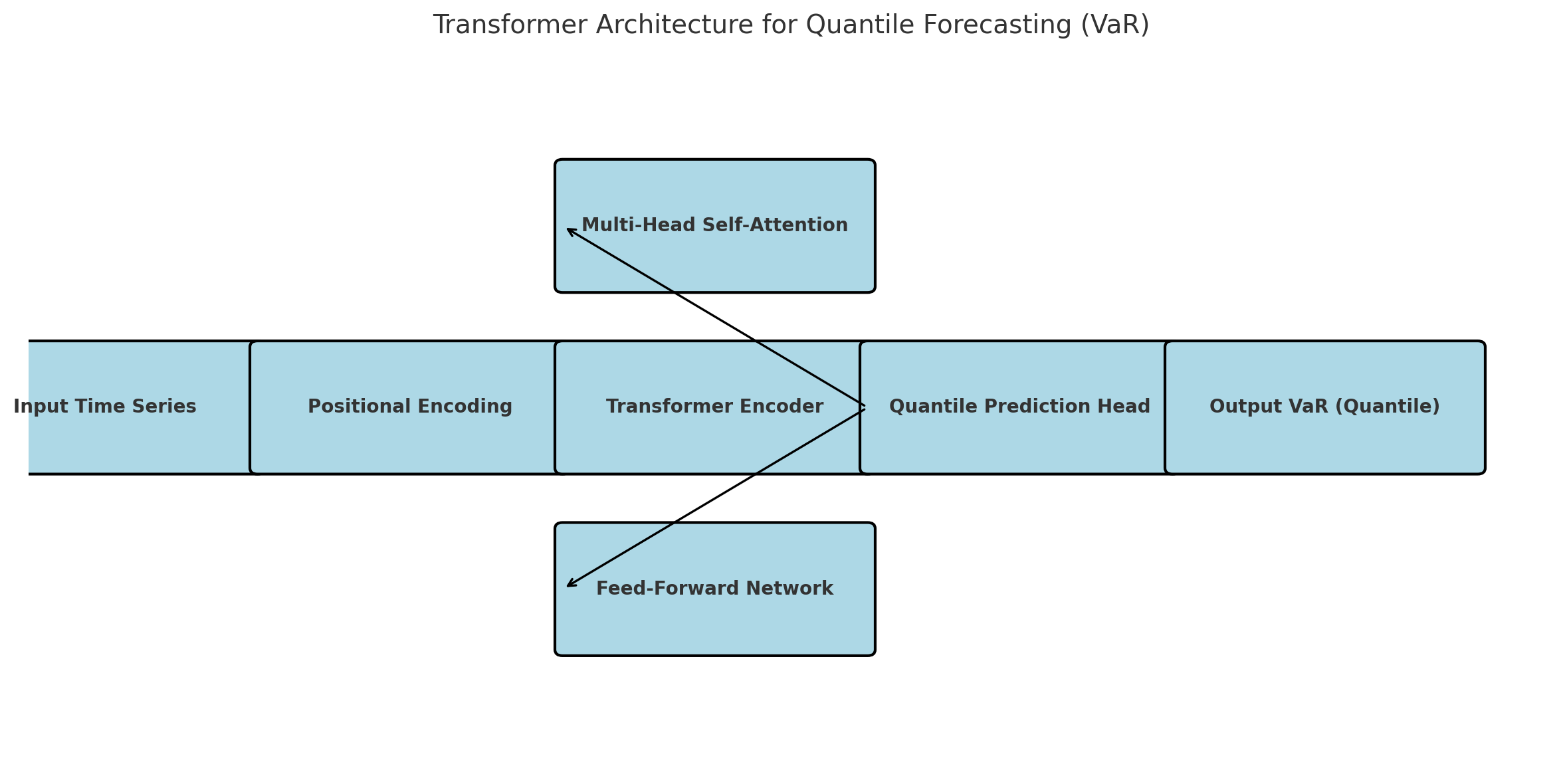}
    \caption{Transformer Architecture for Conditional Quantile Forecasting in VaR Estimation}
    \label{fig:transformer_var}
\end{figure}

Integrating transformer-based quantile regressors into the current framework would enhance its capacity to capture abrupt structural breaks, nonlinear interactions, and endogenous feedback loops—features that are often missed by static or parametric models. Future research may also consider hybrid models that combine deep learning with econometric priors (e.g., GARCH-inspired inductive biases) for improved interpretability and robustness.
Figure~\ref{fig:transformer_var} illustrates the architecture of a Transformer-based model tailored for Value-at-Risk (VaR) estimation via conditional quantile forecasting. The model ingests a multivariate time series input, including historical returns, volatility proxies, macroeconomic indicators, or sentiment indices, which are embedded with positional encodings to retain temporal ordering.

The central module is a Transformer encoder, composed of stacked layers of multi-head self-attention and position-wise feed-forward networks. The self-attention mechanism enables the model to dynamically assign importance to different lags and features, effectively capturing long-range dependencies and regime shifts in financial time series. This contrasts with traditional models such as GARCH, which typically assume fixed lag structures and limited memory depth.

The output of the encoder is passed to a quantile prediction head—typically a dense layer trained with an asymmetric quantile loss function—yielding estimates for specific quantiles of the conditional return distribution (e.g., the 1

This attention-based architecture not only removes the need for strong parametric assumptions (e.g., Gaussianity or fixed volatility dynamics) but also enhances the model's ability to adapt to nonlinear patterns, heavy tails, and abrupt structural breaks. As such, Transformer-based quantile forecasting represents a state-of-the-art approach in data-driven financial risk modeling, offering a robust alternative to traditional econometric methods.

\subsection*{A.2.2 Mathematical Formulation of Transformer Layers}

The Transformer encoder consists of a sequence of layers, each comprising a multi-head self-attention mechanism and a position-wise feed-forward network. These components can be formally described as follows:

\paragraph{Positional Encoding:}  
To encode temporal order, a positional vector $\mathbf{p}_t \in \mathbb{R}^d$ is added to each input embedding $\mathbf{x}_t$. For a time step $t$ and dimension $d$, sinusoidal positional encoding is defined as:

\[
p_t^{(2i)} = \sin\left( \frac{t}{10000^{2i/d}} \right), \quad
p_t^{(2i+1)} = \cos\left( \frac{t}{10000^{2i/d}} \right)
\]

The input to the model becomes $\mathbf{h}_t = \mathbf{x}_t + \mathbf{p}_t$.

\paragraph{Scaled Dot-Product Attention:}  
Given a query $Q$, key $K$, and value $V$ matrices derived from the input embeddings via learned linear projections, the attention mechanism computes:

\[
\text{Attention}(Q, K, V) = \text{softmax}\left( \frac{QK^\top}{\sqrt{d_k}} \right)V
\]

where $d_k$ is the dimension of the key vectors. This allows the model to weigh the importance of different positions in the sequence when predicting each output.

\paragraph{Multi-Head Attention:}  
Instead of performing a single attention function, multi-head attention runs $h$ parallel attention operations:

\[
\text{MultiHead}(Q,K,V) = \text{Concat}(\text{head}_1, \dots, \text{head}_h)W^O
\]

where each head is defined as:

\[
\text{head}_i = \text{Attention}(QW_i^Q, KW_i^K, VW_i^V)
\]

with learnable projection matrices $W_i^Q$, $W_i^K$, $W_i^V$, and $W^O$.

\paragraph{Position-Wise Feed-Forward Network:}  
Each output of the attention layer passes through a two-layer feed-forward network:

\[
\text{FFN}(x) = \max(0, xW_1 + b_1)W_2 + b_2
\]

which is applied identically and independently to each position.

\paragraph{Quantile Output Layer:}  
The final output is passed through a dense layer to generate conditional quantile estimates. For a quantile level $\tau \in (0,1)$, the model is trained to minimize the pinball loss:

\[
\mathcal{L}_\tau(\hat{y}_t, y_t) = \begin{cases}
\tau (y_t - \hat{y}_t), \& y_t > \hat{y}_t \\\\
(1 - \tau)(\hat{y}_t - y_t), \& y_t \leq \hat{y}_t
\end{cases}
\]

This output directly yields $\text{VaR}^{(\tau)}_t = \hat{y}_t$.

\subsection*{A3. A Variance-Decomposition-Based Unified Framework for Systemic Risk}

Diebold and Yılmaz (2014) propose a unified approach to systemic risk quantification based on forecast error variance decompositions (FEVD), which forms the foundation of their connectedness index. Specifically, in a vector autoregressive (VAR) model of order $p$:

\[
\mathbf{y}_t = \sum_{i=1}^{p} \Phi_i \mathbf{y}_{t-i} + \varepsilon_t, \quad \varepsilon_t \sim \mathcal{N}(0, \Sigma)
\]

the $H$-step-ahead forecast error variance decomposition is defined via the moving average representation:

\[
\mathbf{y}_t = \sum_{h=0}^{\infty} \Psi_h \varepsilon_{t-h}
\]

The key metric used to quantify spillovers is the **generalized forecast error variance decomposition (GFEVD)** introduced by Pesaran and Shin (1998), which avoids orthogonalization and ordering issues. The $(j,k)$-th element of the GFEVD matrix is:

\[
\theta_{jk}^{(H)} = \frac{\sigma_{kk}^{-1} \sum_{h=0}^{H-1} \left( \mathbf{e}_j' \Psi_h \Sigma \mathbf{e}_k \right)^2 }{ \sum_{h=0}^{H-1} \left( \mathbf{e}_j' \Psi_h \Sigma \Psi_h' \mathbf{e}_j \right) }
\]

where:
- $\sigma_{kk}$ is the $k$-th diagonal element of $\Sigma$
- $\mathbf{e}_j$ is a selection vector (1 in $j$-th position, 0 elsewhere)
- $\Psi_h$ are the moving average coefficients
- $H$ is the horizon length

These quantities are normalized row-wise to ensure that:

\[
\tilde{\theta}_{jk}^{(H)} = \frac{\theta_{jk}^{(H)}}{\sum_{k=1}^{N} \theta_{jk}^{(H)}}
\quad \text{such that} \quad \sum_{k=1}^{N} \tilde{\theta}_{jk}^{(H)} = 1
\]

Based on this, the connectedness measures are constructed as follows:

- Total Connectedness Index (TCI):
\[
\mathcal{C}^{(H)} = \frac{1}{N} \sum_{j=1}^{N} \sum_{\substack{k=1 \\ k \neq j}}^{N} \tilde{\theta}_{jk}^{(H)} \times 100
\]

- Directional “To” Connectedness from $j$ to others:
\[
\mathcal{C}_{j \rightarrow \cdot}^{(H)} = \sum_{\substack{k=1 \\ k \neq j}}^{N} \tilde{\theta}_{kj}^{(H)} \times 100
\]

- Directional “From” Connectedness received by $j$:
\[
\mathcal{C}_{\cdot \rightarrow j}^{(H)} = \sum_{\substack{k=1 \\ k \neq j}}^{N} \tilde{\theta}_{jk}^{(H)} \times 100
\]

- Net Directional Connectedness:
\[
\mathcal{C}_{j}^{\text{net},(H)} = \mathcal{C}_{j \rightarrow \cdot}^{(H)} - \mathcal{C}_{\cdot \rightarrow j}^{(H)}
\]

\textbf{Interpretation:}
- The “from” degree $\mathcal{C}_{\cdot \rightarrow j}^{(H)}$ measures how much variable $j$ is influenced by others — similar in spirit to Marginal Expected Shortfall (MES).
- The “to” degree $\mathcal{C}_{j \rightarrow \cdot}^{(H)}$ reflects how much systemic impact $j$ transmits — akin to $\Delta$CoVaR.
- The total connectedness $\mathcal{C}^{(H)}$ summarizes global fragility or stress in the financial system.
\begin{figure}[H]
    \centering
    \includegraphics[width=0.4\textwidth]{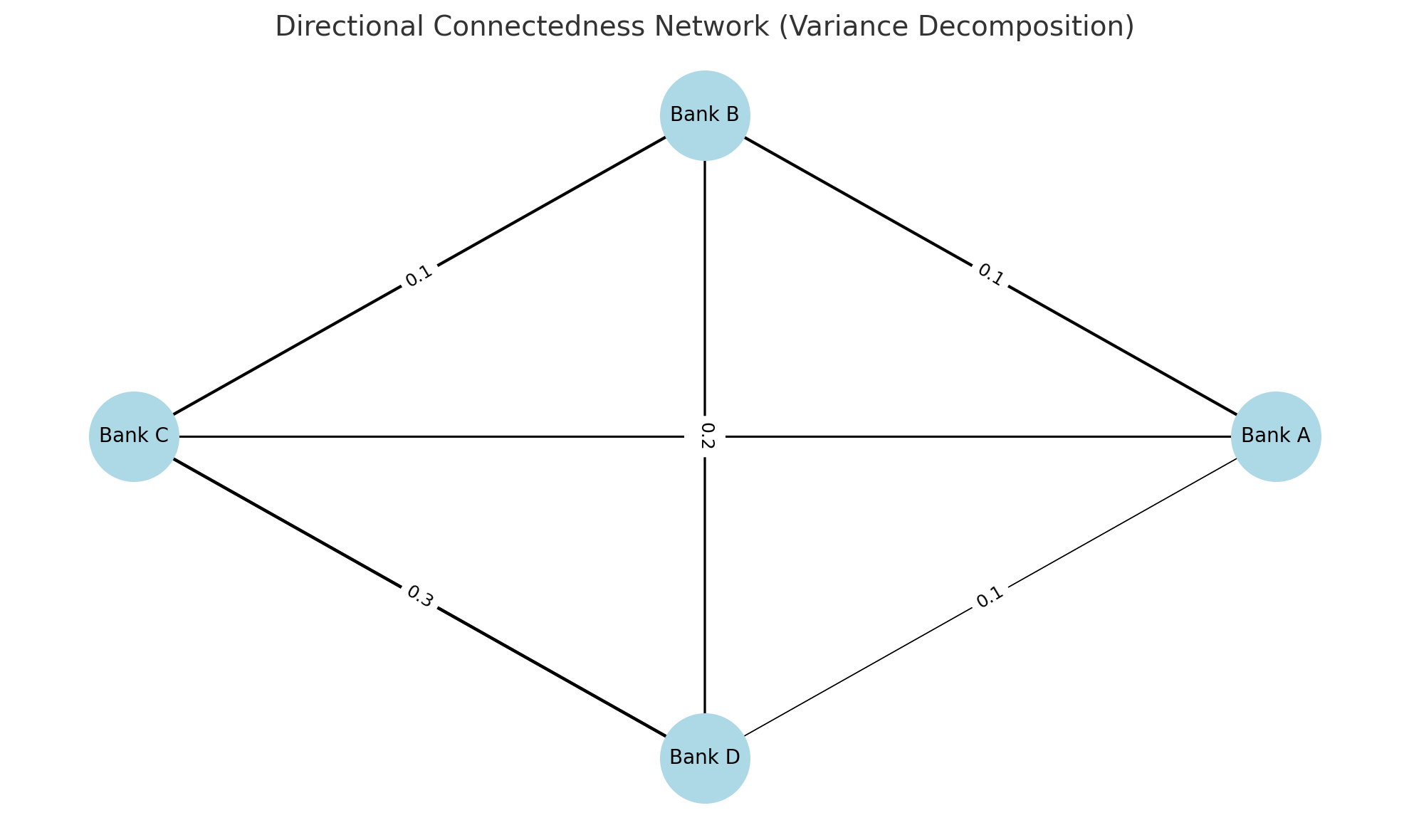}
    \caption{Directional Connectedness Network Constructed from Variance Decomposition}
    \label{fig:connectedness_network}
\end{figure}

\textbf{Policy Implication:} These connectedness measures provide dynamic, real-time proxies for systemic risk and can be integrated with tail risk models (e.g., VaR backtesting, Expected Shortfall) to form a coherent, variance-based systemic risk monitoring toolkit.

\subsection*{Conclusion}

By incorporating formal statistical testing procedures, adopting state-of-the-art deep learning models for tail risk estimation, and embedding risk metrics within a generalized variance decomposition network framework, future research can establish a more unified, dynamic, and empirically grounded approach to financial risk measurement. Such enhancements are not only theoretically justified but are also increasingly relevant for systemic risk surveillance in high-frequency, interdependent financial environments.

\section*{References}

\begin{enumerate}
    \item Alexander, C. (2008). \textit{Market Risk Analysis, Volume IV: Value-at-Risk Models}. John Wiley \& Sons.

    \item Jorion, P. (2007). \textit{Value at Risk: The New Benchmark for Managing Financial Risk} (3rd ed.). McGraw-Hill.

    \item Engle, R. F. (1982). Autoregressive Conditional Heteroskedasticity with Estimates of the Variance of United Kingdom Inflation. \textit{Econometrica}, 50(4), 987–1007. \url{https://doi.org/10.2307/1912773}

    \item Bollerslev, T. (1986). Generalized Autoregressive Conditional Heteroskedasticity. \textit{Journal of Econometrics}, 31(3), 307–327.

    \item Barone-Adesi, G., Giannopoulos, K., \& Vosper, L. (1999). VaR without correlations for portfolios of derivative securities. \textit{Journal of Futures Markets}, 19(5), 583–602.

    \item Yamai, Y., \& Yoshiba, T. (2005). Value-at-Risk versus Expected Shortfall: A Practical Perspective. \textit{Journal of Banking \& Finance}, 29(4), 997–1015.

    \item Basel Committee on Banking Supervision. (2016). \textit{Minimum Capital Requirements for Market Risk}. Bank for International Settlements.

    \item OpenAI. (2024). \textit{ChatGPT Model (gpt-4)}. Accessed via OpenAI API and ChatGPT Web Platform. \url{https://openai.com/chatgpt}
    
    \item Diebold, F. X., \& Yılmaz, K. (2014). On the network topology of variance decompositions: Measuring the connectedness of financial firms. \textit{Journal of Econometrics}, 182(1), 119–134. \url{https://doi.org/10.1016/j.jeconom.2014.04.012}
\end{enumerate}



\end{document}